\documentclass[intlimits,twoside,a4paper]{article}

\usepackage[cp1251]{inputenc}

\usepackage[eqsecnum]{cmpj3}
\usepackage{bm}
\issue{2020}{23}{4}{43705}
\doinumber{10.5488/CMP.23.43705}
\title[Spin-1/2 $XXZ$ Heisenberg cupolae]%
{Spin-1/2 $XXZ$ Heisenberg cupolae: magnetization process and related enhanced magnetocaloric effect}
\author[K. Kar\v{l}ov\'a]{K. Kar\v{l}ov\'a}
\address{
 Institute of Physics, Faculty of Science, P.J. \v{S}af\'{a}rik University, Park Angelinum 9, 040 01 Ko\v{s}ice, Slovak Republic}

\date{Received June 23, 2020, in final form July 30, 2020}

\begin{document}

\maketitle

\begin{abstract}
The magnetization curves and magnetocaloric effect during the adiabatic demagnetization of  antiferromagnetic spin-1/2 $XXZ$ Heisenberg clusters with the shape of Johnson's solids (triangular cupola, square cupola and pentagonal cupola) are investigated using the exact numerical diagonalization  for different values of the exchange anisotropy ranging in between the Ising and fully isotropic limit. It is demonstrated that spin-1/2 $XXZ$ Heisenberg cupolae display, in comparison with their Ising counterparts, at least one more magnetization plateau. The novel magnetization plateaux extends over a wider range of the magnetic fields with increasing of a quantum part of the $XXZ$ exchange coupling at the expense of the original plateaux present in the  limiting Ising case. It is shown that the spin-1/2  $XXZ$ Heisenberg triangular and pentagonal cupolae exhibit an enhanced magnetocaloric effect in the vicinity of zero magnetic field, which  makes these magnetic systems promising refrigerants for cooling down to ultra-low zero temperatures.
\end{abstract}

\section{Introduction}
Magnetic systems composed of a finite number of interacting spins have attracted considerable attention due to the possibility of exploring many fundamental physical phenomena \cite{frie18,hoty18}. Antiferromagnetic small spin clusters can exhibit plethora features in the zero-temperature magnetization curve, for instance magnetization steps and jumps \cite{shap02,schu02}, magnetization plateaux \cite{lacr11,hone04} or quasi-plateau \cite{bell14,ohan15}. Intermediate magnetization plateaux were predicted for  the Ising spin clusters with the shape of Platonic solids \cite{stre15}, `star of David' \cite{zuko18} or tetrahedra-based Ising clusters \cite{mohy19}, while the magnetization curve of Heisenberg spin clusters was studied with the shape of triangle \cite{hara05}, tetrahedron \cite{bose05}, truncated tetrahedron \cite{schn09,coff92}, cuboctahedron \cite{rous08,karl17}, icosidodecahedron \cite{rous08} and many more \cite{hzad19,kons05,kons09,stre14,szal20,hzad20}.  It is noteworthy that besides the theoretical predictions, the magnetization plateaux were indeed observed in experimental representatives of molecular magnets, for example for linear trimer compound A$_3$Cu(PO$_4$)$_4$ \cite{beli05}, cubane-based compounds \cite{aron07}, or homodinuclear nickel complexes \cite{stre05,stre08}.

In addition, antiferromagnetic spin clusters open the way towards several applications. Among them there could be mentioned the magnetic cooling based on the magnetocaloric effect \cite{evan06,sess12,evan14,zhen14}, which is especially pronounced in frustrated quantum spin systems \cite{zhit03}. The magnetocaloric effect has attracted much attention in the study of low-dimensional magnetic systems both from theoretical \cite{topi12,derz06,rich06,derz04,schn13} and experimental \cite{hoop16,hoop12} points of view due to the fact that an enhanced magnetocaloric effect represents the basis for economically cheaper and environmentally friendly refrigeration technology compared to standard refrigeration technology based on vapor expansion \cite{tish03}. An enhanced magnetocaloric effect was found in the vicinity of magnetization jumps, around which the rapid change of temperature was observed upon the variation of the magnetic field during the process of adiabatic demagnetization. Moreover, it was shown in our recent study \cite{karlo17} that the absence of zero magnetization plateau causes an enhanced magnetocaloric effect when shutting down the external magnetic field during the process of adiabatic demagnetization. It was found that the Ising octahedron, dodecahedron and cuboctahedron exhibit this feature and they are promising candidates for reaching ultra-low temperatures \cite{karl17,karlo17}. Unfortunately, by considering the ($xy$) part of the exchange interaction, the temperature achieves only finite values by switching-off  the magnetic field during the adiabatic demagnetization due to the presence of zero magnetization plateau in the magnetization process of the spin-1/2 $XXZ$ Heisenberg octahedron, dodecahedron and cuboctahedron. The Heisenberg spin clusters with the absence of zero magnetization plateau as  promising candidates for future applications in magnetic refrigeration has been much less studied. Therefore, it appears worthwhile to investigate the low-temperature magnetization process and isentropes in the field-temperature plane of the spin-1/2 $XXZ$ Heisenberg cupolae, for two of which we are able to predict the absence of zero magnetization plateau in the zero-temperature magnetization curve. It should be pointed out that  magnetization occurs in the magnetization curve of spin-1/2 clusters at the following fractional values
\begin{eqnarray}
\frac{m}{m_\text{s}}=\frac{\frac{N}{2}-n}{\frac{N}{2}}\,,
\label{rule}
\end{eqnarray}
where $N$ is the total number of spins of a given cluster, $N/2$ denotes the maximum value for $z$-component of the total spin $S_\text{T}^z$, and $n=0,1,\ldots,\lfloor N/2 \rfloor$ determines deviation of the $z$-component of the total spin $S_\text{T}^z$ from its maximum value. 
It is clear from equation (\ref{rule}) that the existence of the zero magnetization plateau is possible only if $N=2n$, i.e., if $N$ is an even number.

In the present paper we focus our attention on the magnetization process and magnetocaloric effect of the spin-1/2 $XXZ$ Heisenberg triangular and pentagonal cupola with an odd number of spins $N=9$ and $N=15$. In addition, we  investigate the spin-1/2 $XXZ$ Heisenberg square cupola with the total number of spins $N=12$, for which  several experimental realizations were found in the magnetic compounds Sr(TiO)Cu$_4$(PO$_4$)$_4$ \cite{kato19,isla18}, Pb(TiO)Cu$_4$(PO$_4$)$_4$ \cite{kimu18}, Ba(TiO)Cu$_4$(PO$_4$)$_4$ \cite{babk17,kato17,kimura18,rast,kimu16}. 

The organization of this paper is as follows. The spin-1/2 $XXZ$ Heisenberg cupolae are introduced in section \ref{model}, where the  method of calculation used is also presented. The most interesting results for the ground-state phase diagram, magnetization process and isentropes in the field-temperature plane are discussed in section \ref{results}. Finally, several concluding remarks are mentioned in section \ref{conclusion}.

\section{Spin-1/2 $XXZ$ Heisenberg cupolae}
\label{model}

\begin{figure}[!b]
\begin{center}
\includegraphics[width=0.8\textwidth]{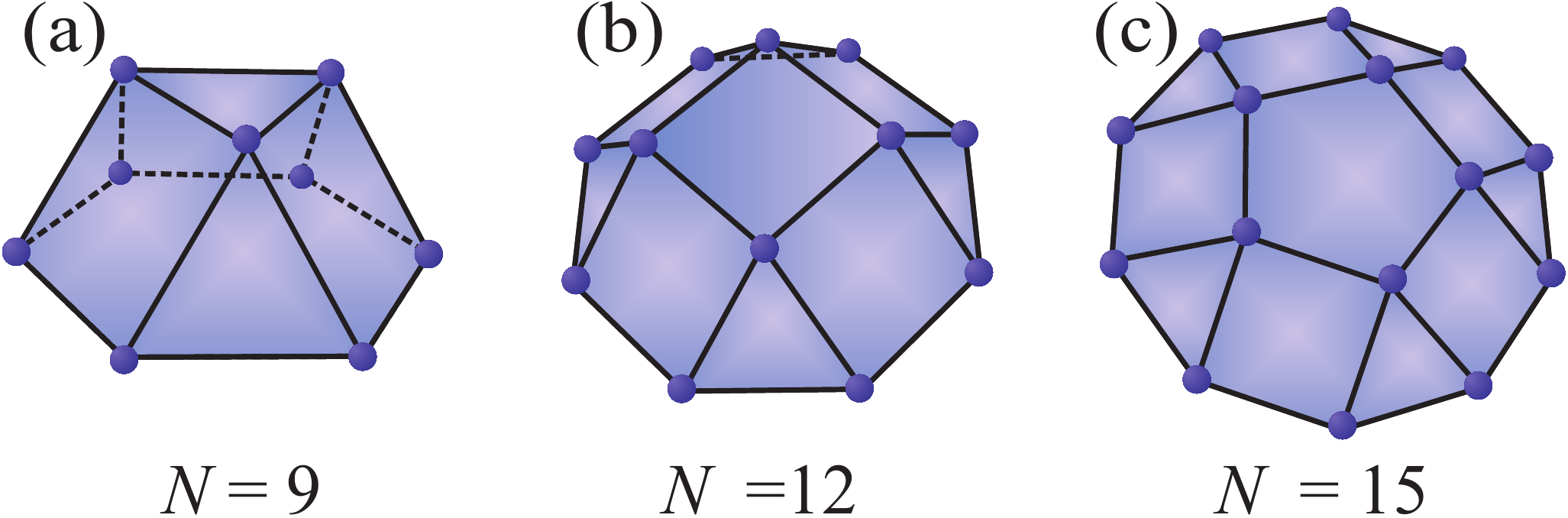}
\end{center}
\caption{(Colour online) A steric arrangement of the spin-1/2 Heisenberg cupolae: (a) triangular cupola; (b) square cupola; (c) pentagonal cupola. The number $N$ determines the total number of spins of a given $XXZ$ Heisenberg cupola.}
\label{cupoly}
\end{figure}
Let us examine the spin-1/2 $XXZ$ Heisenberg clusters with the shape of Johnson's cupolae (see figure~\ref{cupoly}), which are defined through the following Hamiltonian
\begin{eqnarray}
\hat{\cal H}=J\sum_{\langle i,j\rangle}\left[\Delta\left(\hat{S}_i^x\hat{S}_j^x + \hat{S}_i^y\hat{S}_j^y\right)+\hat{S}_i^z\hat{S}_j^z\right]-h\sum_{i=1}^N\hat{S}_i^z,
\label{ham}
\end{eqnarray}
where $\hat{S}_i^{\alpha}$ denotes the spatial projections ($\alpha=x,y,z$) of the spin-1/2 operator placed at $i$-th site of Johnson cupolae, the first summation takes into account the exchange interaction of antiferromagnetically coupled    nearest-neighbour spins ($J>0$) and $\Delta \in \langle 0;1\rangle$ represents the anisotropy parameter in the $XXZ$ exchange interaction, i.e., $\Delta=0$ corresponds to the Ising model and $\Delta =1$ corresponds to the isotropic Heisenberg model. The second term in the Hamiltonian (\ref{ham}) expresses the Zeeman energy of magnetic moments in nonzero magnetic field $h>0$, $N$ denotes the total number of spins. Exact diagonalization data for the isothermal magnetization curves and entropy of the spin-1/2 $XXZ$ Heisenberg cupolae were obtained by the use of the subroutine fulldiag from the Algorithms and Libraries for Physics Simulations (ALPS) project \cite{bau11}. This technique allows us to gain magnetization curves and magnetocaloric properties of the spin-1/2 $XXZ$ Heisenberg cupolae.

\section{Results and discussion}
\label{results}
\begin{figure}[!b]
\begin{center}
\includegraphics[width=0.48\textwidth]{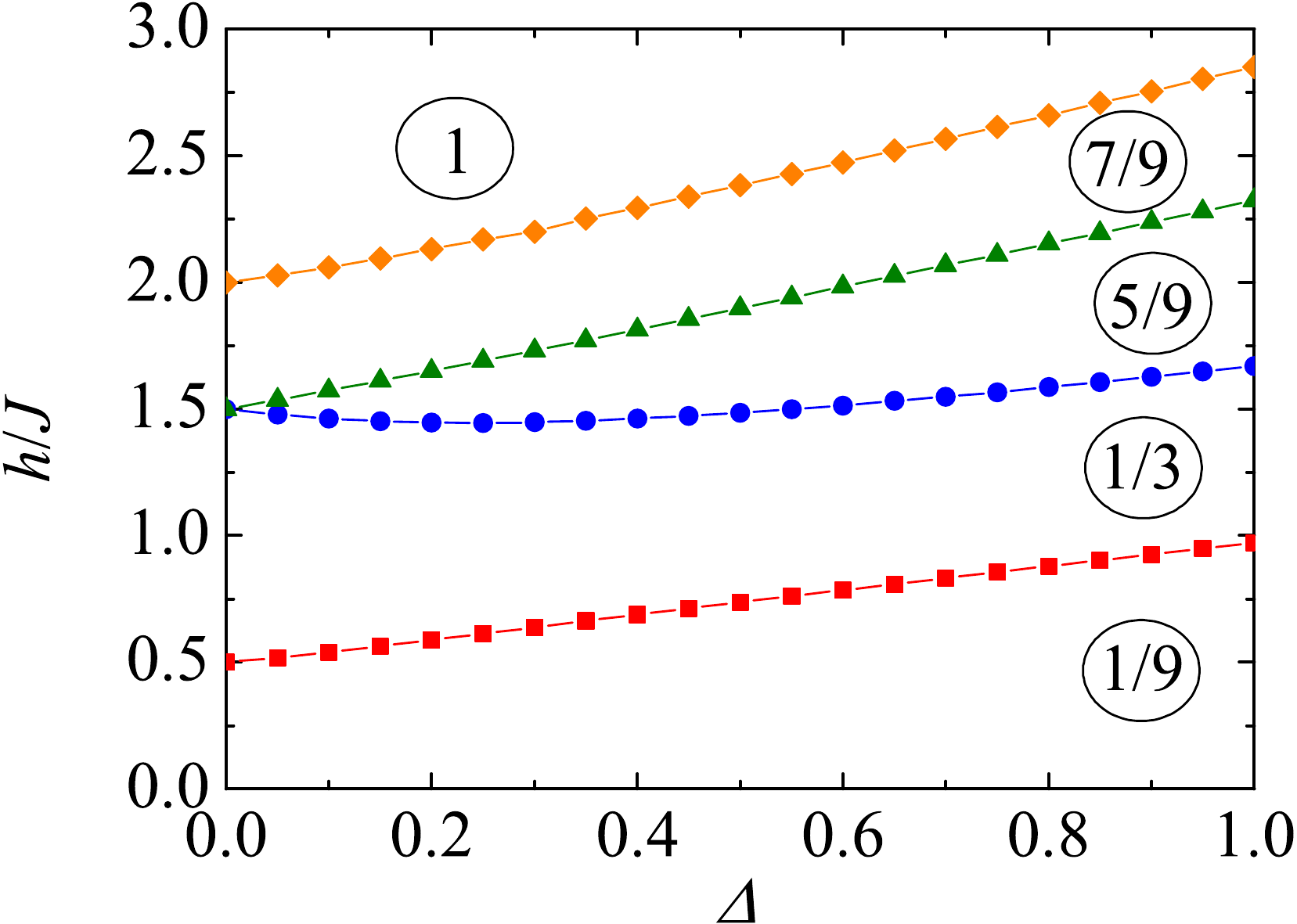}
\end{center}
\caption{(Colour online) The ground-state phase diagram of the spin-$1/2$ $XXZ$ Heisenberg triangular cupola in the $\Delta{-}h/J$ plane. Each delimited area represents lowest-energy eigenstate with the magnetization normalized with respect to its saturation values as quoted by numbers given in circles.
}
\label{tGSPD}
\end{figure}
Let us start our discussion with the ground-state phase diagram of the spin-1/2 $XXZ$ Heisenberg triangular cupola, which is depicted in figure \ref{tGSPD} in the $\Delta{-}h/J$ plane. There are five distinct phases in the ground-state phase diagram of the $XXZ$ Heisenberg triangular cupola for any $\Delta \neq 0$, which are delimited by displayed level-crossing fields and which differ by the value of the total magnetization normalized with respect to its saturation value. Contrary to this, five-ninth plateau is missing in the ground-state phase diagram of the spin-1/2 Ising triangular cupola ($\Delta=0$).

\begin{figure}[!b]
\includegraphics[width=0.49\textwidth]{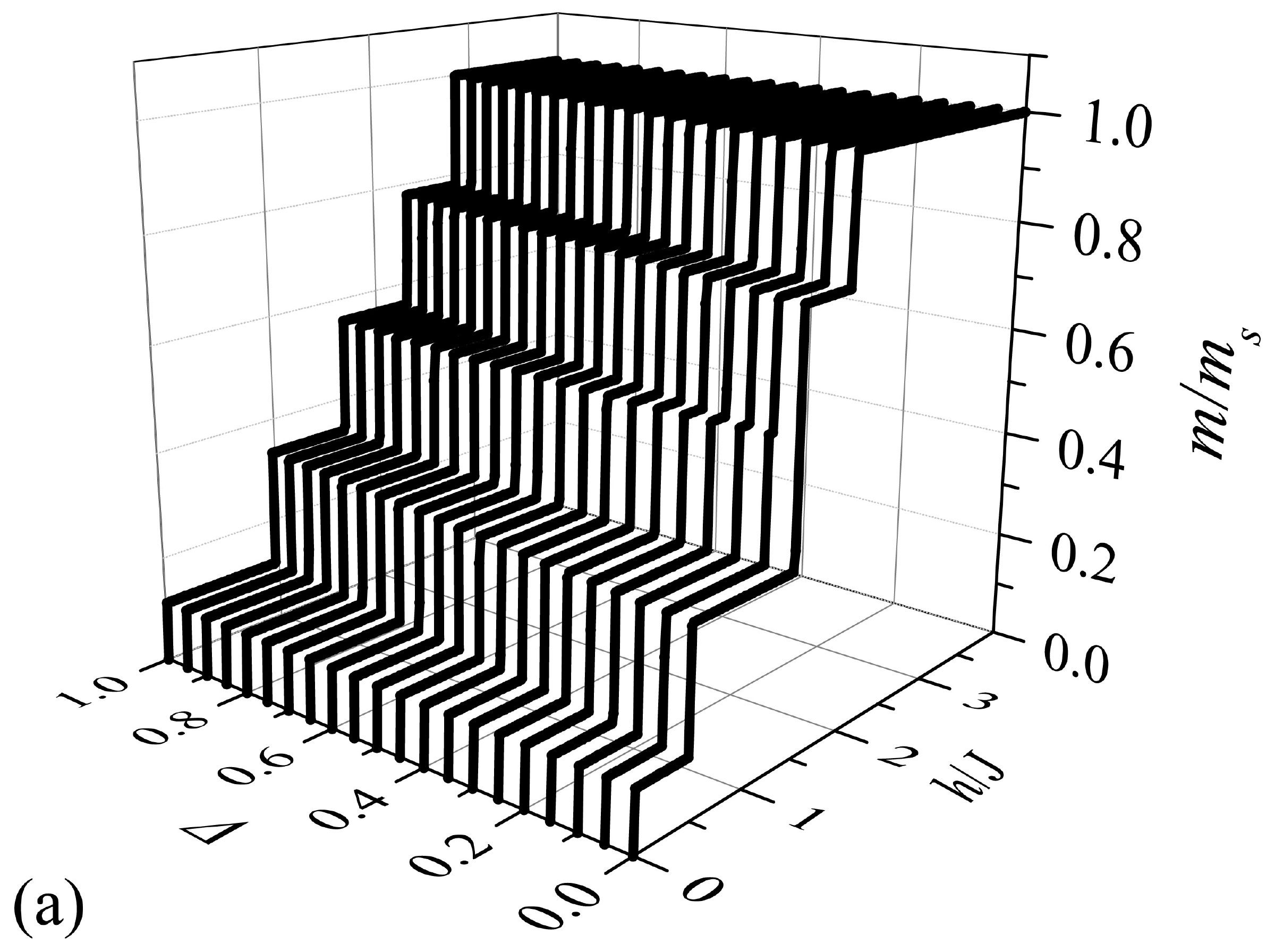}
\includegraphics[width=0.49\textwidth]{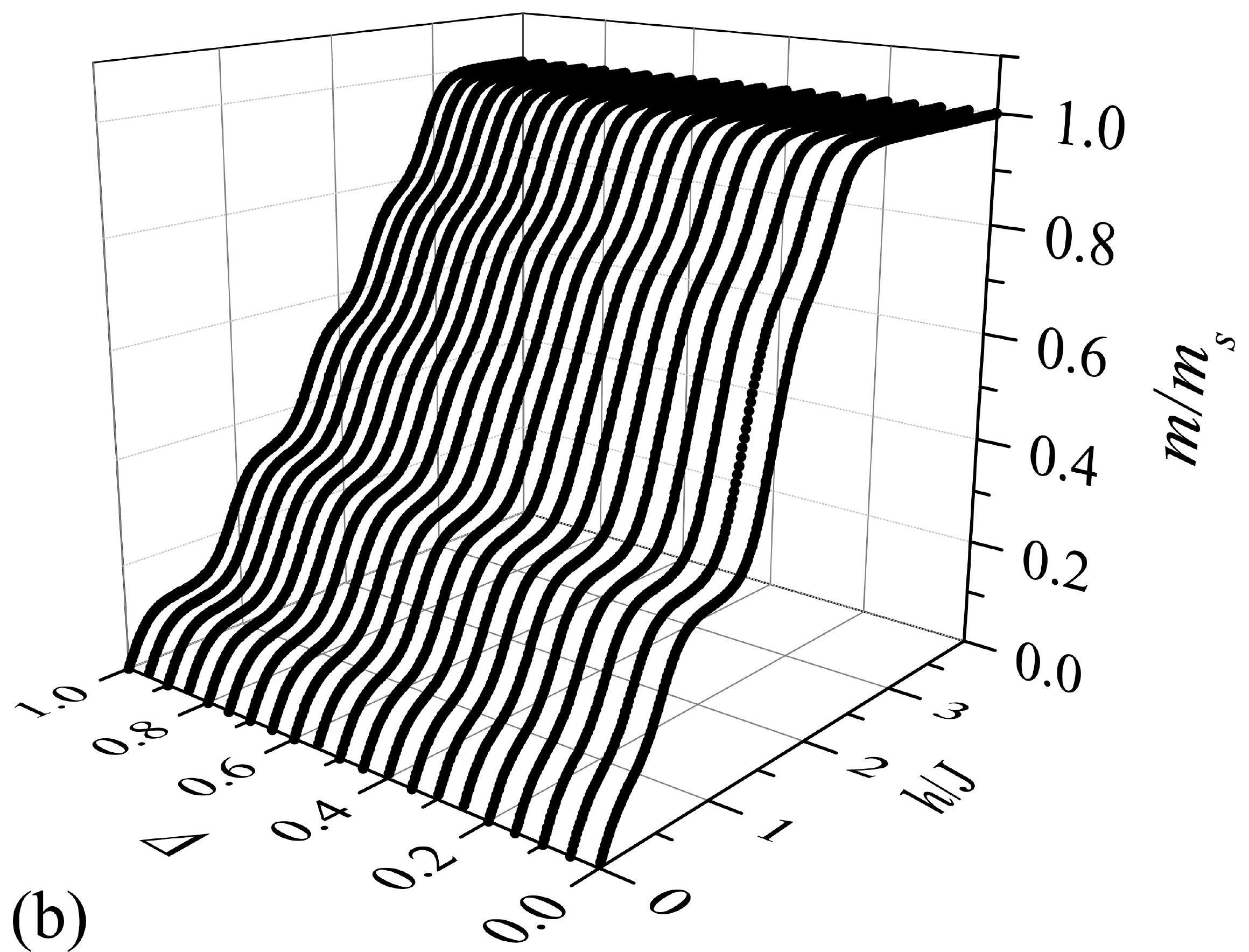}
\caption{The magnetization curves of a spin-1/2 $XXZ$ Heisenberg triangular cupola for several values of the anisotropy parameter $\Delta$ and two different temperatures: (a) $k_{\textrm B}T/J=0.001$; (b) $k_{\textrm B}T/J=0.1$.}
\label{tmag}
\end{figure}
To better understand how the low-energy eigenstates influence the magnetization process at finite temperatures, we depict in figure \ref{tmag} the field dependence of the isothermal magnetization for several values of the anisotropy parameter $\Delta$ at two different temperatures $k_{\textrm B}T/J=0.001$ and $0.1$. It can be noticed in figure~\ref{tmag}~(a)  that the low-temperature magnetization curve at $k_{\textrm B}T/J=0.001$ strongly resembles zero-temperature variations of the magnetizations, showing almost abrupt magnetization jumps at the level-crossing fields.  As one can see in figure \ref{tmag}~(a), the low-temperature magnetization curve of spin-1/2 Ising triangular cupola indeed   exhibits, beside saturation value,  three intermediate plateaux at one-ninth, one-third and seven-ninth of the saturation magnetization. In addition, one more five-ninth plateau can be observed in a low-temperature magnetization curve of the spin-1/2 $XXZ$ Heisenberg triangular cupola for any $\Delta\neq 0$, which extends over a wider range of magnetic fields with increasing of the anisotropy parameter $\Delta$. The magnetization curves of the spin-1/2 $XXZ$ Heisenberg cupola at the moderate temperature $k_{\textrm B}T/J=0.1$ are depicted in figure \ref{tmag}~(b). As expected, magnetization curves and jumps are gradually smoother upon increasing the temperature. It can be seen in figure \ref{tmag}~(b) that all magnetization plateaux except the widest ones are reflected by inflection points only.

\begin{figure}[!b]
\vspace{-0.3cm}
\includegraphics[width=0.49\textwidth]{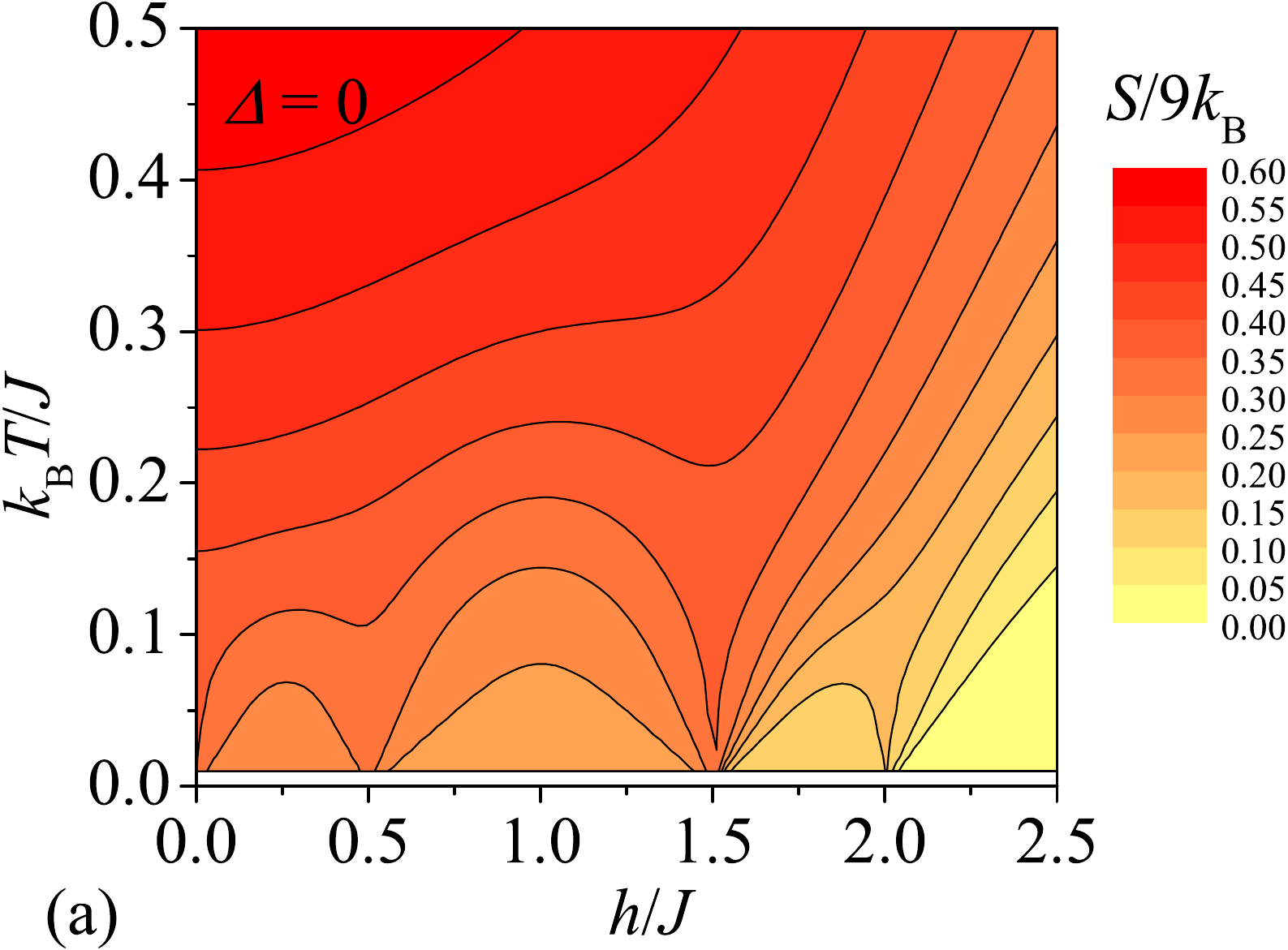}
\includegraphics[width=0.49\textwidth]{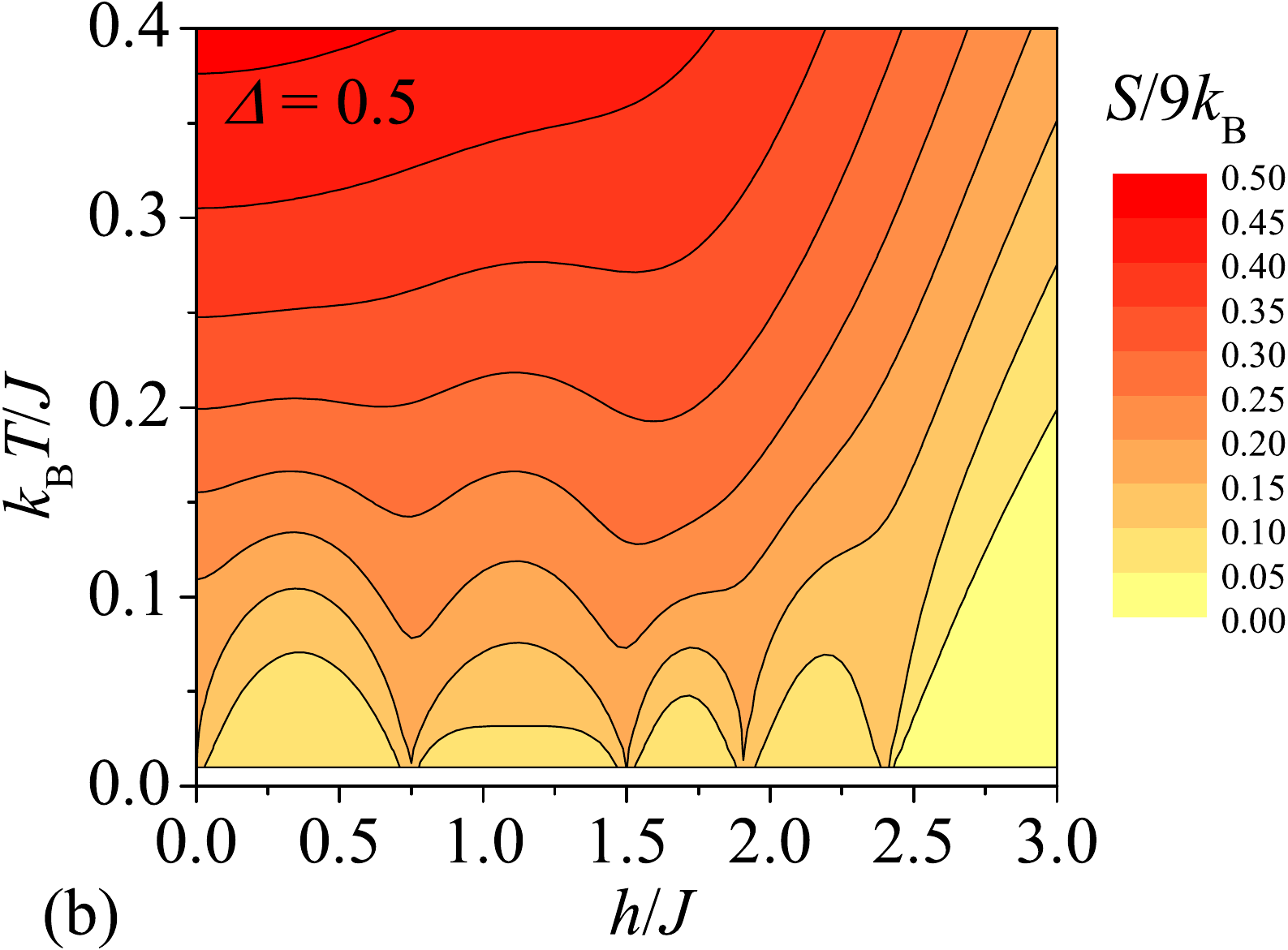}
\begin{center}
\includegraphics[width=0.49\textwidth]{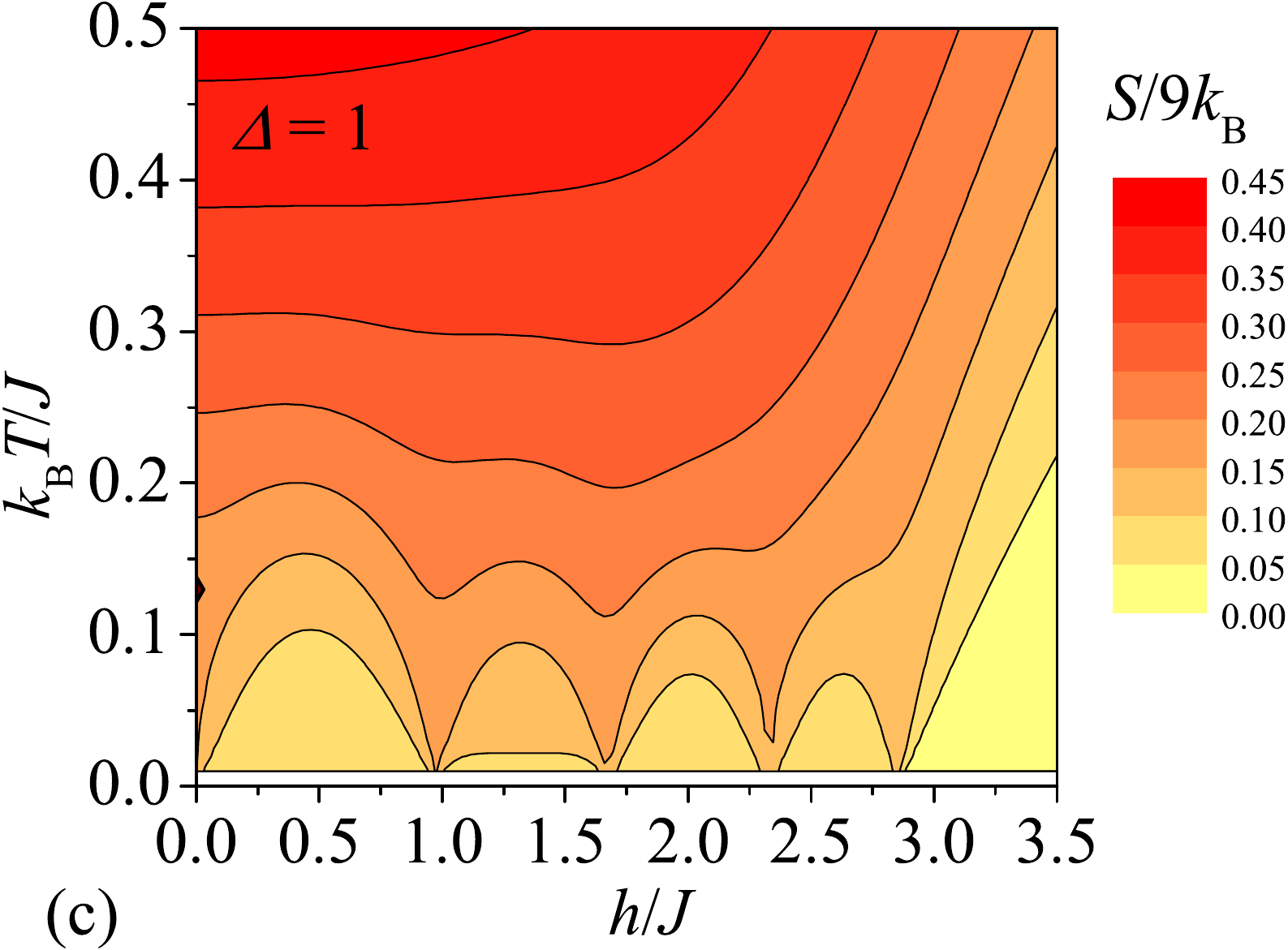}
\end{center}
\vspace{-0.3cm}
\caption{(Colour online) The isentropes of a spin-1/2 $XXZ$ Heisenberg triangular cupola in the $h/J{-}k_{\textrm {B}}T/J$ plane for three different values of the anisotropy parameter: (a) $\Delta=0$; (b) $\Delta=0.5$; (c) $\Delta=1$. Bottom horizontal lines correspond to the lowest temperature $k_{\textrm{B}}T/J=0.01$ used for numerical calculations.}
\label{tdem}
\end{figure}
Next, let us also examine the isentropes of the spin-1/2 $XXZ$ Heisenberg triangular cupola, which can be especially fascinating in proximity of the magnetization jumps. A few isentropes are displayed in figure~\ref{tdem}~(a) for the spin-1/2 Ising triangular cupola in the $h/J{-}k_{\textrm B}T/J$ plane. The plotted isentropes determine a magnetocaloric effect of the spin-1/2 $XXZ$ Heisenberg triangular cupola, in particular, the adiabatic temperature change with the change of external magnetic field. 
The temperature of the Ising triangular cupola sharply varies in the vicinity of all critical fields, in which magnetization jumps occur. Owing to this fact,  the Ising triangular cupola exhibits a direct magnetocaloric effect just above the level-crossing fields, while  an inverse one is detected just below the level-crossing fields. It should be noted that a giant magnetocaloric effect in the proximity of zero magnetic field is potentially applicable for achieving ultra-low temperatures quite similarly to the Ising octahedron, dodecahedron and cuboctahedron \cite{karlo17,karl17}. Furthermore, the existence of an enhanced magnetocaloric effect,  from the technological point of view, is most valuable near the zero magnetic field, which is most frequently used during the process of adiabatic demagnetization as a final value of the external magnetic field.

An adiabatic demagnetization of the $XXZ$ Heisenberg triangular cupola with relatively strong anisotropy parameter $\Delta=0.5$ is displayed in figure \ref{tdem}~(b), which shows a contour plot of entropy as a function of temperature and magnetic field. A giant magnetocaloric effect can be found in the proximity of all magnetization jumps of the spin-1/2 $XXZ$ Heisenberg triangular cupola [see figures \ref{tdem}~(a) and~\ref{tdem}~(b)]. Moreover, the absence of zero-magnetization plateau for all values of anisotropy parameter $\Delta$, brings about the possibility of effective cooling during the  adiabatic demagnetization. This is in contrast to the spin-1/2 $XXZ$ Heisenberg octahedron, dodecahedron and cuboctahedron \cite{karlo17,karl17}, which exhibit a giant magnetocaloric effect when switching-off  the external magnetic field only for the Ising limiting case $\Delta=0$. A few isentropes of the spin-1/2 isotropic Heisenberg triangular cupola ($\Delta=1$) are plotted in figure \ref{tdem}~(c). As one can see in figure \ref{tdem}~(c), the density plot of the entropy of isotropic Heisenberg triangular cupola is in qualitative concordance with the $XXZ$ triangular cupola at the relatively strong anisotropy parameter $\Delta=0.5$.

\begin{figure}[!b]
\begin{center}
\includegraphics[width=0.48\textwidth]{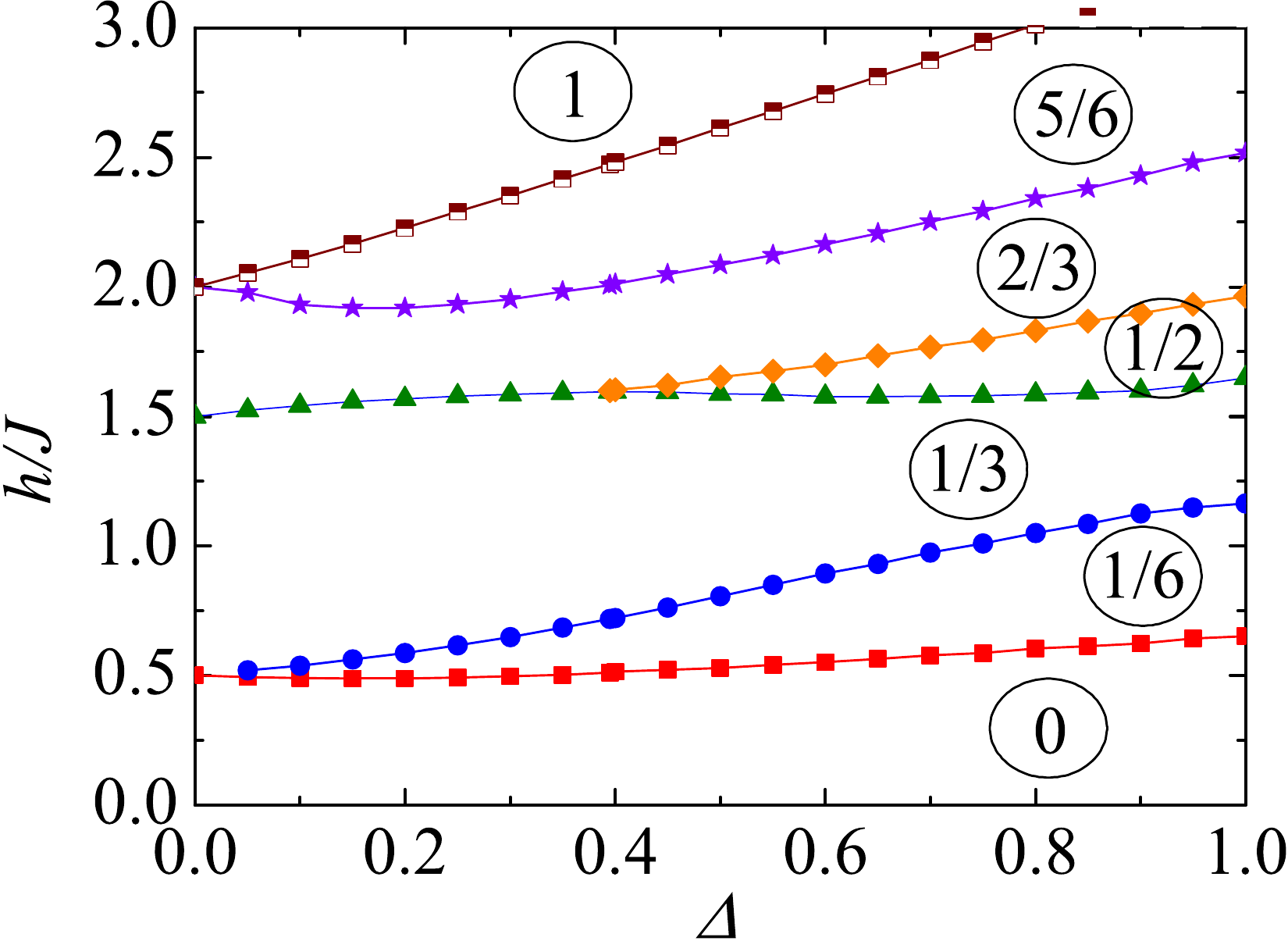}
\end{center}
\vspace{-.3cm}
\caption{(Colour online) The ground-state phase diagram of the spin-$1/2$ $XXZ$ Heisenberg square cupola in the $\Delta{-}h/J$ plane. Each delimited area represents lowest-energy eigenstate with the magnetization normalized with respect to its saturation values as quoted by numbers given in circles.}
\label{fig5}
\end{figure}
The ground-state phase diagram of the spin-1/2 $XXZ$ Heisenberg square cupola is shown in figure \ref{fig5} in the $\Delta{-}h/J$ plane. The Ising square cupola ($\Delta=0$) exhibits three intermediate plateaux at zero, one-third and two-thirds of the saturation magnetization in zero-temperature magnetization process. When introducing $xy$-part of the $XXZ$ exchange interaction $\Delta>0$, two additional magnetization plateaux  evolve in a wider interval of magnetic fields. In addition, zero-temperature magnetization process of the $XXZ$ Heisenberg square cupola exhibits one extra plateau if the anisotropy parameter reaches the value $\Delta\approx 0.395$.

\begin{figure}[!t]
\includegraphics[width=0.49\textwidth]{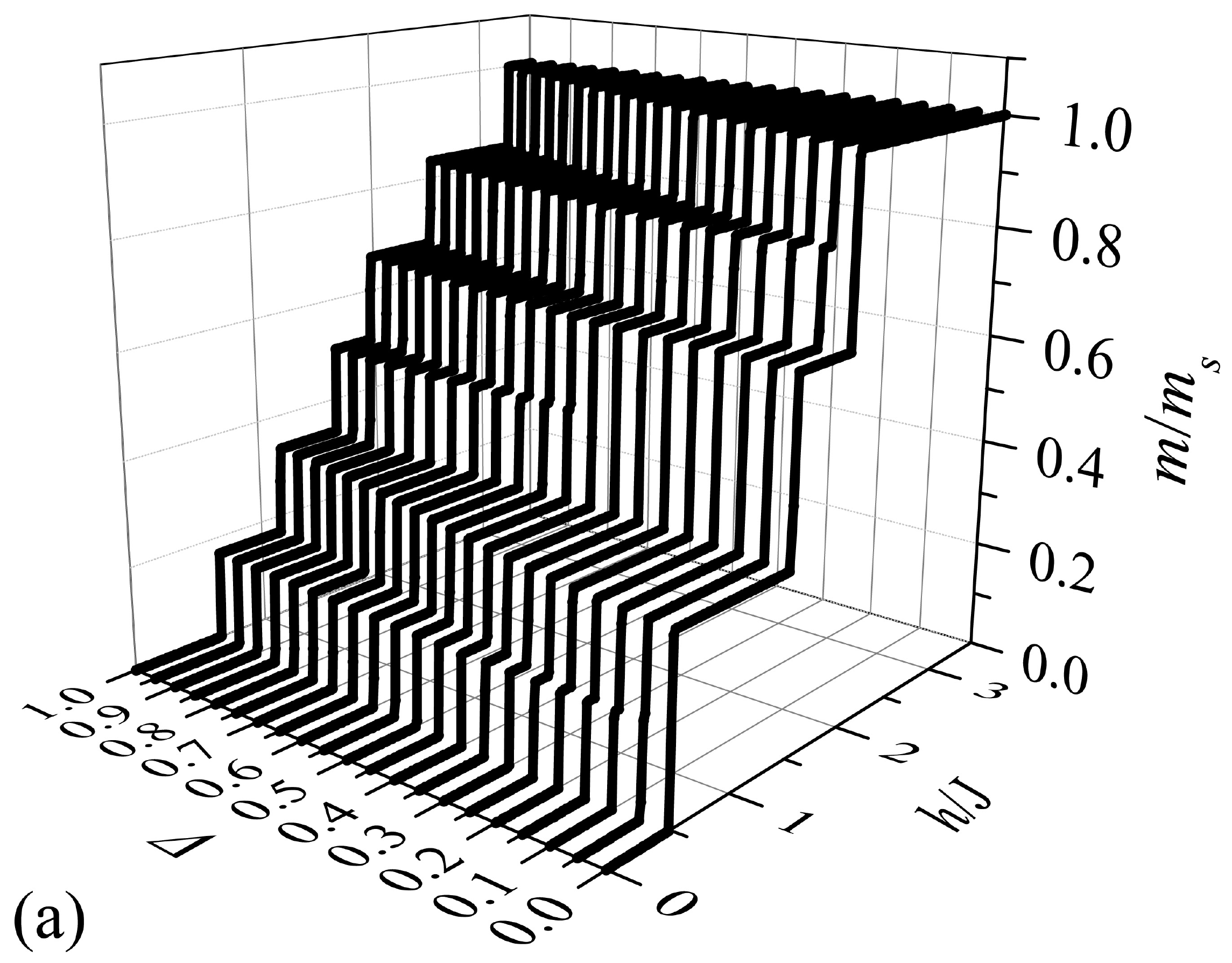}
\includegraphics[width=0.49\textwidth]{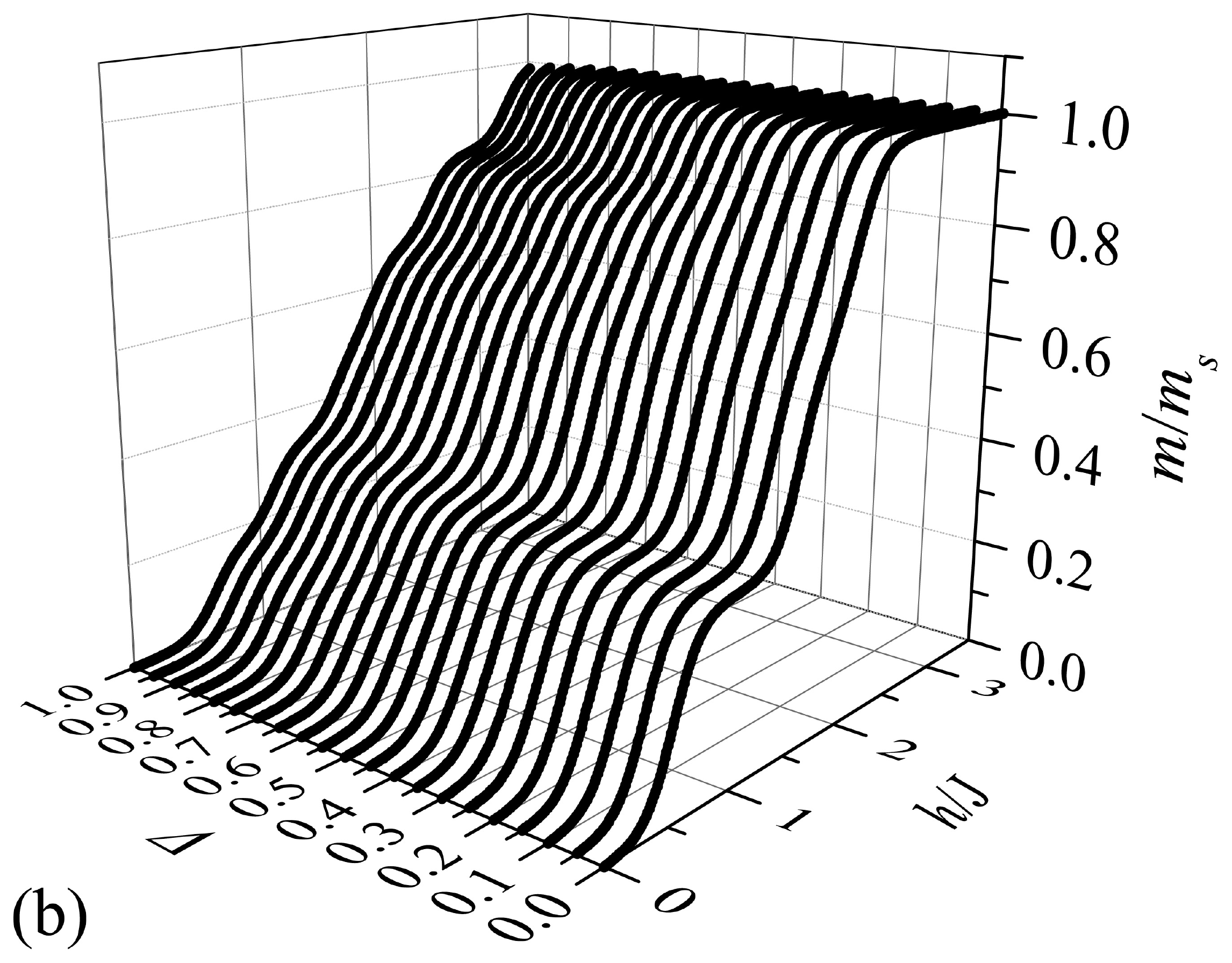}
\caption{Magnetization curves of a spin-1/2 $XXZ$ Heisenberg square cupola for several values of the anisotropy parameter $\Delta$ and two different temperatures: (a) $k_{\textrm B}T/J=0.001$; (b) $k_{\textrm B}T/J=0.1$.}
\label{fig6}
\end{figure}
\begin{figure}[!t]
	\includegraphics[width=0.49\textwidth]{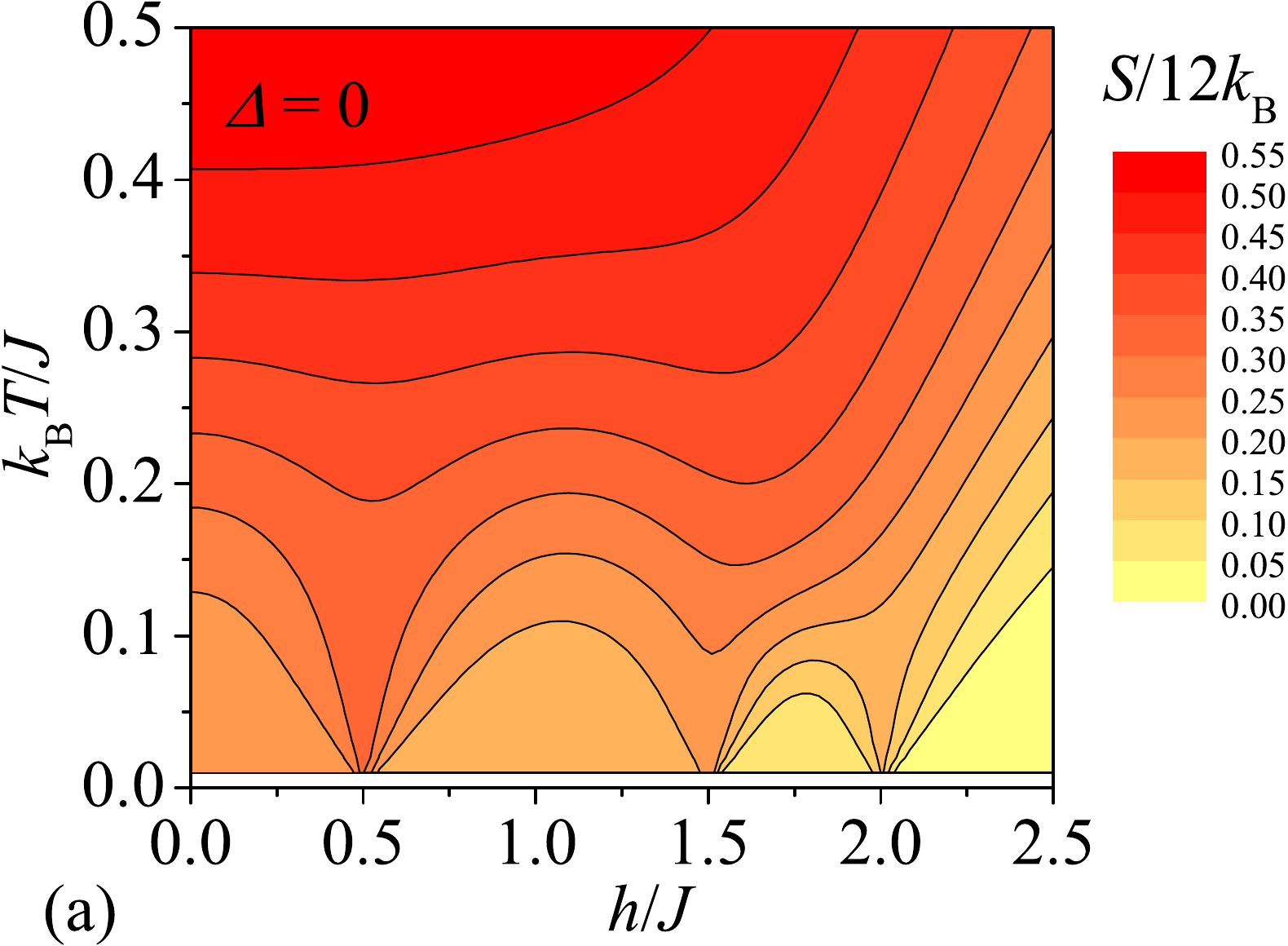}
	\includegraphics[width=0.49\textwidth]{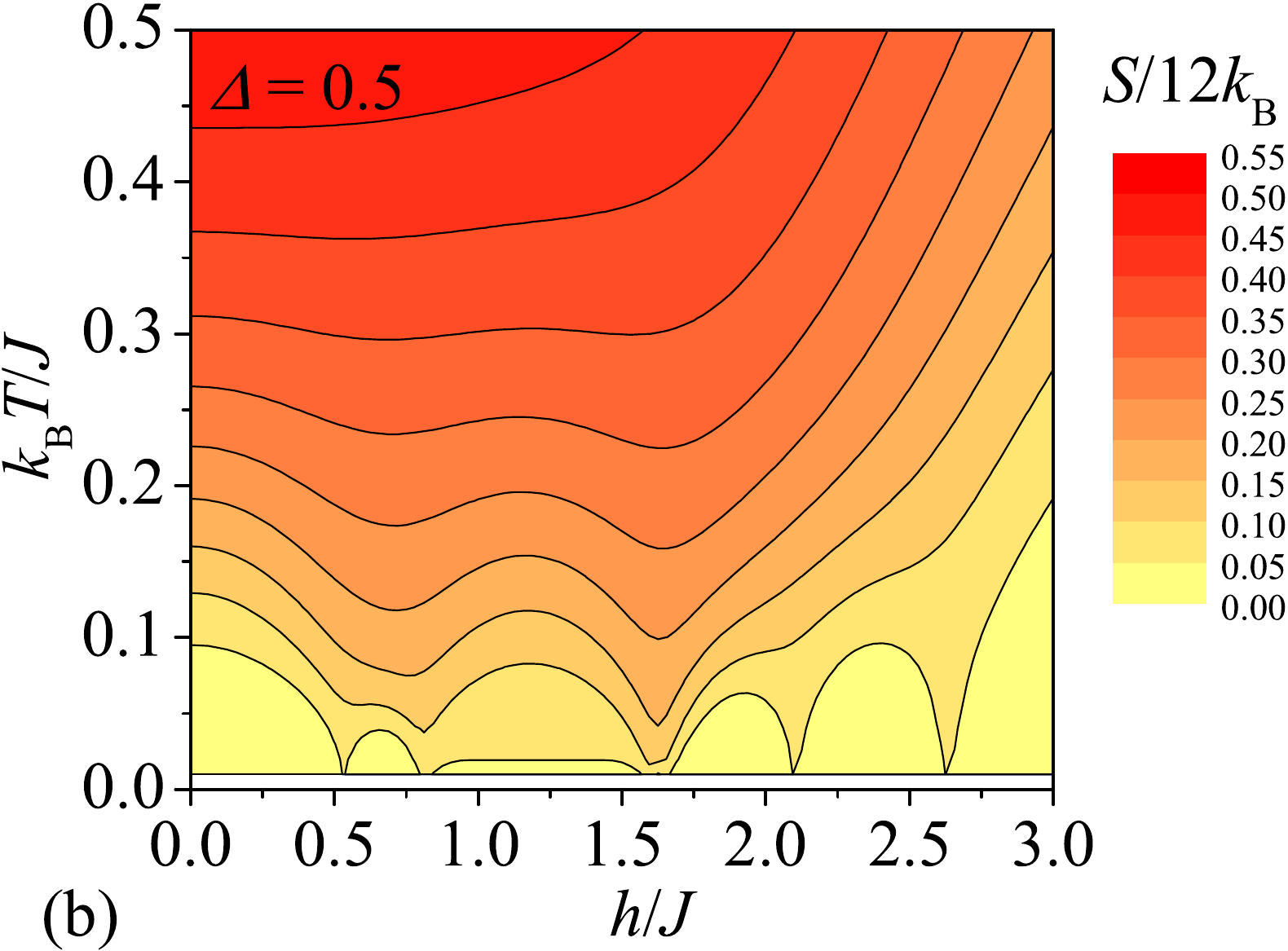}
	\begin{center}
		\includegraphics[width=0.5\textwidth]{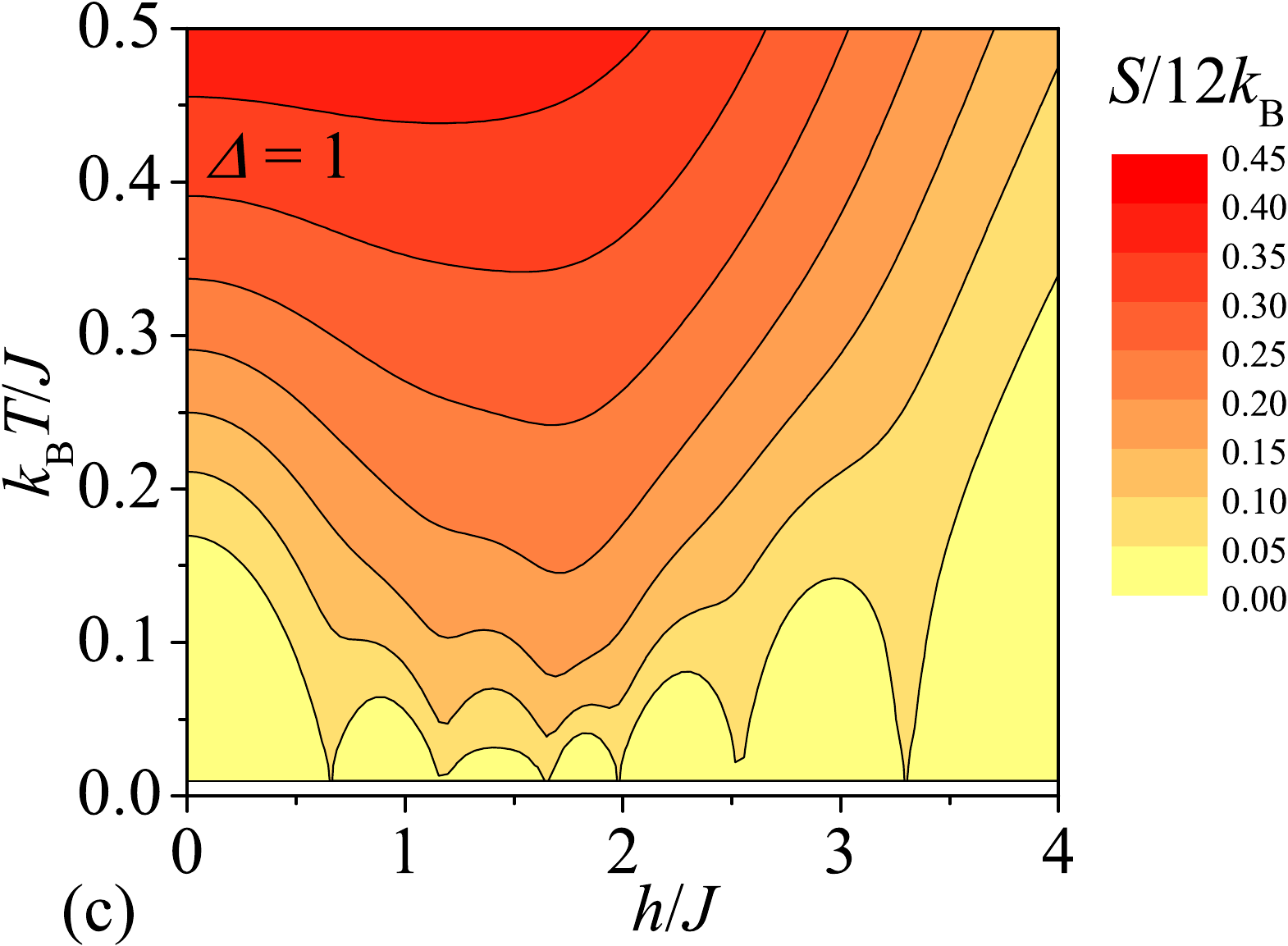}
	\end{center}
	\caption{(Colour online) Isentropes of a spin-1/2 $XXZ$ Heisenberg square cupola in the $h/J{-}k_{\textrm {B}}T/J$ plane for three different values of the anisotropy parameter: (a) $\Delta=0$; (b) $\Delta=0.5$; (c) $\Delta=1$. Bottom horizontal lines correspond to the lowest temperature $k_{\textrm {B}}T/J=0.01$ used for numerical calculations.}
	\label{fig7}
\end{figure}

To confirm this statement, we have depicted in figure \ref{fig6} the isothermal magnetization curves of a spin-1/2 $XXZ$ Heisenberg square cupola for several values of the anisotropy parameter $\Delta$ and two different temperatures. As one can see in figure \ref{fig6}~(a), the low-temperature magnetization curve indeed exhibits three magnetization plateaux if $\Delta=0$ and totally five (six) intermediate plateaux if $0<\Delta<0.395$ ($\Delta>0.395$). The influence of higher temperature on the magnetization process of the spin-1/2 $XXZ$ Heisenberg square cupola is shown in figure \ref{fig6}~(b). At a moderate temperature $k_{\textrm B}T/J=0.1$, only those intermediate plateaux are still visible which were found in low-temperature magnetization curve in a relatively wide interval of magnetic fields.
Furthermore, it is clear from figure \ref{fig7} that the magnetocaloric response of the spin-1/2 $XXZ$ Heisenberg square cupola depends on the choice of the exchange anisotropy~$\Delta$. If $\Delta=0$ is set, the spin-1/2 Ising square cupola exhibits a steep change of temperature around three different magnetic fields. Contrary to this,  the spin-1/2 $XXZ$ Heisenberg square cupola shows an enhanced magnetocaloric effect in the proximity of five (six) level-crossing fields for $\Delta=0.5$ ($\Delta=1$).

\begin{figure}[!t]
\begin{center}
\includegraphics[width=0.48\textwidth]{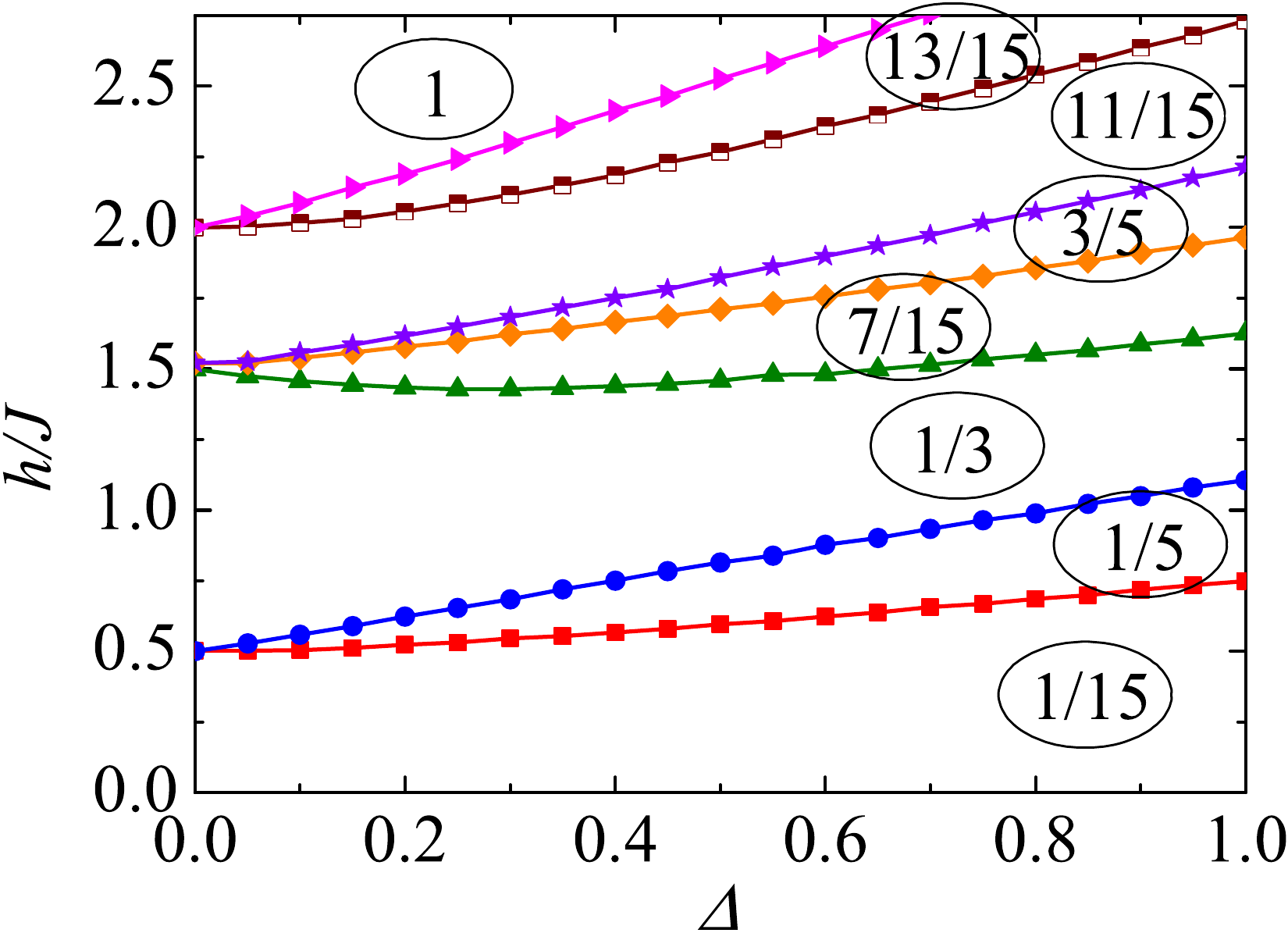}
\end{center}
\caption{(Colour online) The ground-state phase diagram of the spin-$1/2$ $XXZ$ Heisenberg pentagonal cupola in the $\Delta{-}h/J$ plane. Each delimited area represents lowest-energy eigenstate with the magnetization normalized with respect to its saturation values as quoted by numbers given in circles.}
\label{fig8}
\end{figure}
\begin{figure}[!t]
\includegraphics[width=0.49\textwidth]{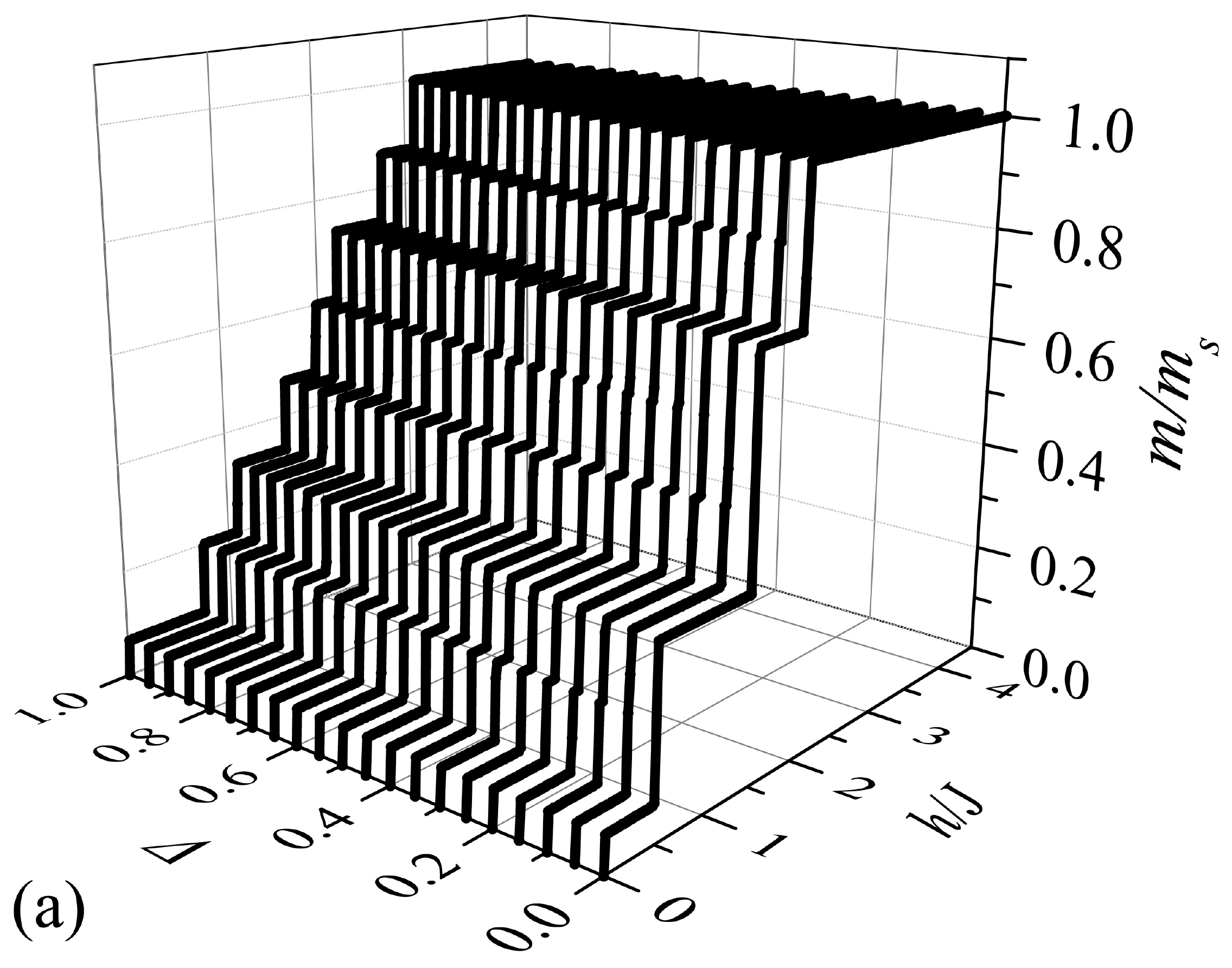}
\includegraphics[width=0.49\textwidth]{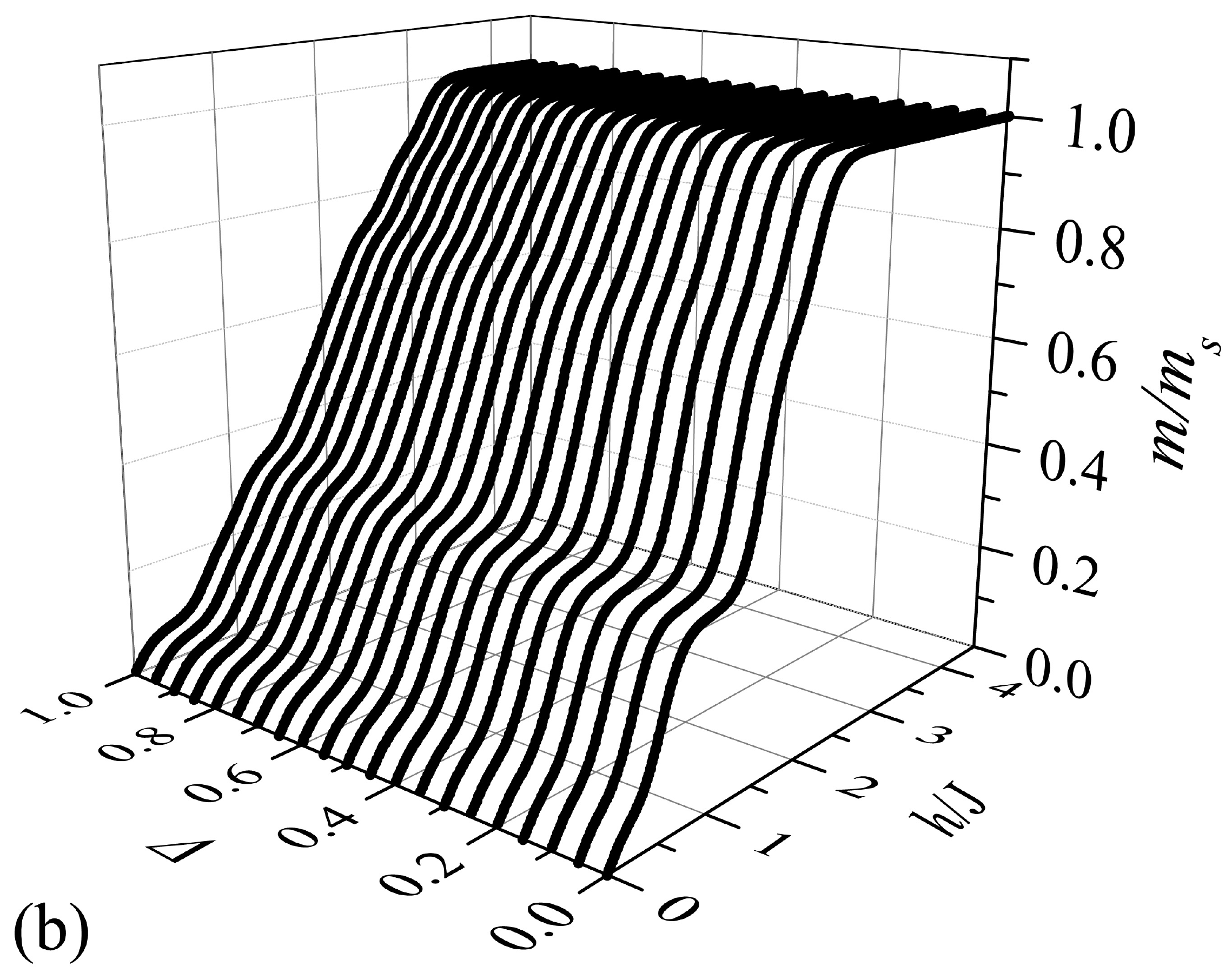}
\caption{The magnetization curves of a spin-1/2 $XXZ$ Heisenberg pentagonal cupola for several values of the anisotropy parameter $\Delta$ and two different temperatures: (a) $k_{\textrm B}T/J=0.001$; (b) $k_{\textrm B}T/J=0.1$.}
\label{fig9}
\end{figure}
Last but not least, the ground-state phase diagram of the spin-1/2 $XXZ$ Heisenberg pentagonal cupola is depicted in figure \ref{fig8} in the $\Delta{-}h/J$ plane. It can be seen in figure \ref{fig8} that the zero-temperature magnetization curve of the spin-1/2 Ising pentagonal cupola ($\Delta=0$) should exhibit three magnetization jumps at level-crossing fields $h/J=0.5$, 1.5 and $2.0$, which determine rise and fall of one-fifteenth, one-third and eleven-fifteenths magnetization plateau. On the other hand, the spin-1/2 $XXZ$ Heisenberg pentagonal cupola with $\Delta\neq 0$ should additionally display four further lowest-energy eigenstates, which correspond  to the one-fifth, seven-fifteenths, three-fifths and thirteen-fifteenths magnetization plateaux in the zero-temperature magnetization curve.

The above mentioned findings can be corroborated by the isothermal magnetization curves, which are depicted in figure \ref{fig9}~(a) for a spin-1/2 $XXZ$ Heisenberg pentagonal cupola at low temperature $k_{\textrm B}T/J=0.001$. As one can see in figure \ref{fig9}~(a), the low-temperature magnetization process of the spin-1/2 Ising pentagonal cupola exhibits three intermediate magnetization plateaux if $\Delta=0$, while the spin-1/2 Heisenberg pentagonal cupola shows seven intermediate magnetization plateaux whenever $\Delta\neq 0$. It can be found from figure \ref{fig9}~(b) that the moderate temperature $k_{\textrm B}T/J=0.1$ is sufficient to suppress all intermediate plateaux except the widest one-fifth and one-third plateau at relatively weak and relatively strong values of the exchange anisotropy $\Delta$, respectively.

\begin{figure}[!t]
\includegraphics[width=0.49\textwidth]{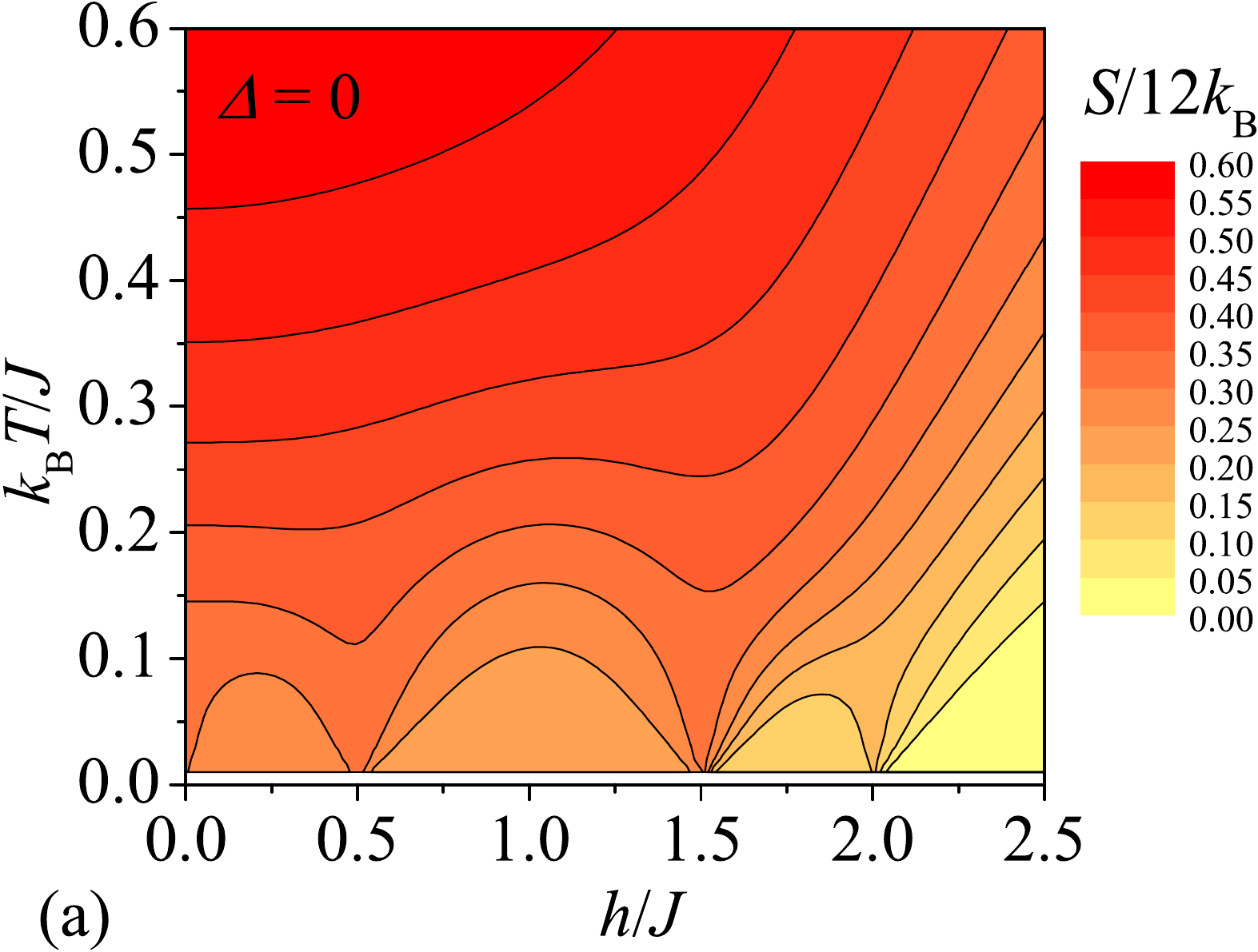}
\includegraphics[width=0.49\textwidth]{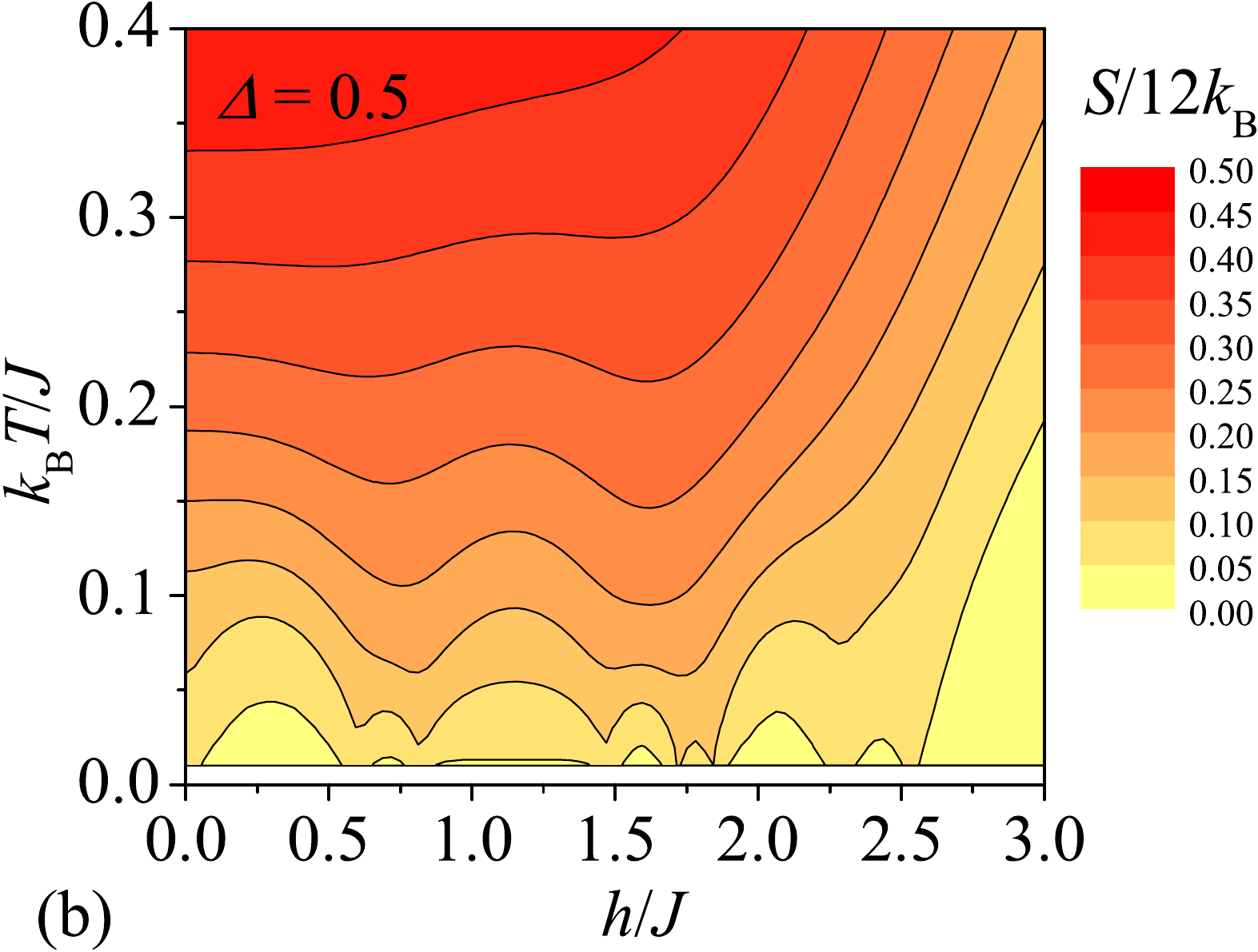}
\begin{center}
\includegraphics[width=0.49\textwidth]{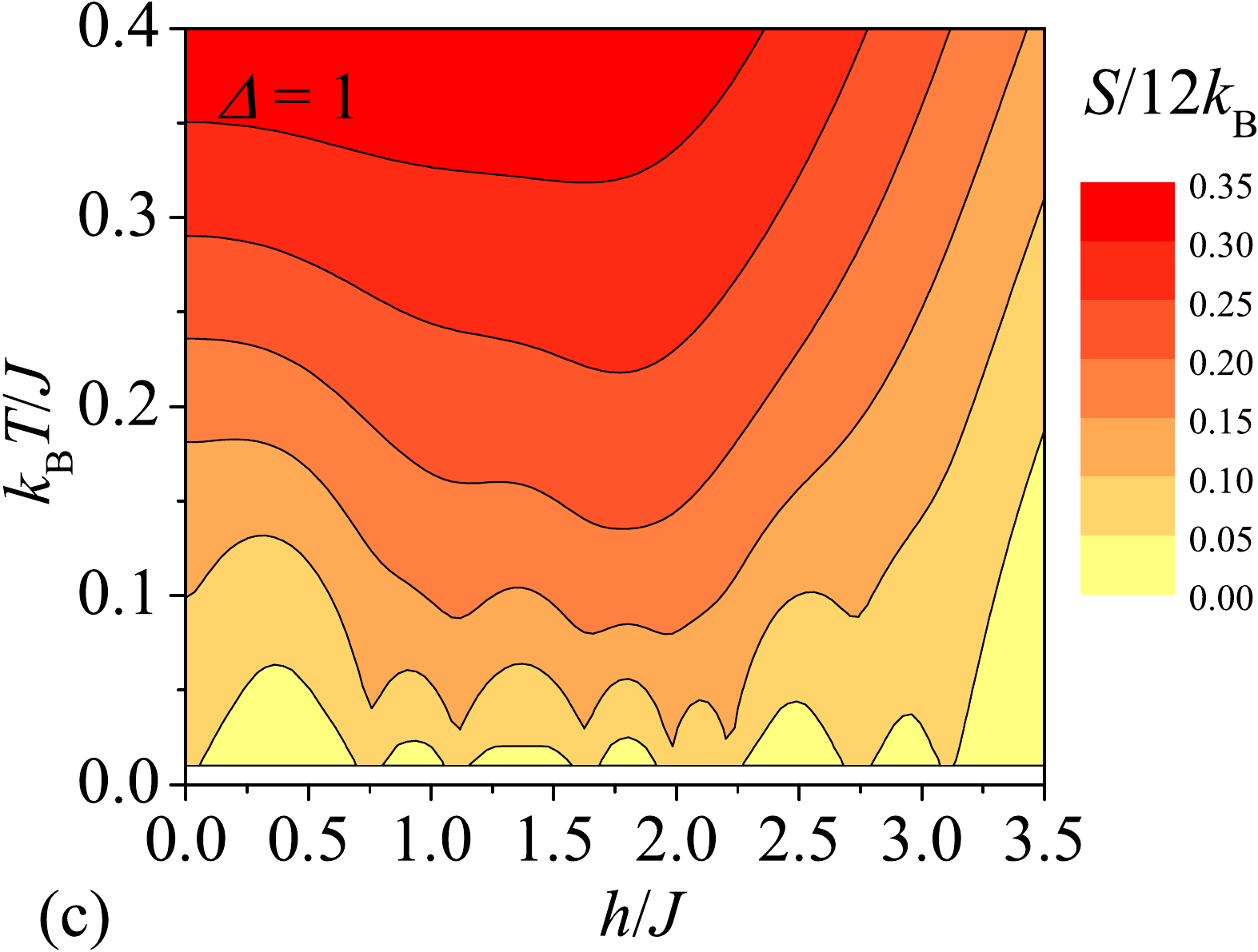}
\end{center}
\caption{(Colour online) The isentropes of a spin-1/2 $XXZ$ Heisenberg pentagonal cupola in the $h/J{-}k_{\textrm {B}}T/J$ plane for three different values of the anisotropy parameter: (a) $\Delta=0$; (b) $\Delta=0.5$; (c) $\Delta=1$. Bottom horizontal lines correspond to the lowest temperature $k_{\textrm {B}}T/J=0.01$ used for numerical calculations.}
\label{fig10}
\end{figure}

For completeness, a few isentropes of the spin-1/2 $XXZ$ Heisenberg pentagonal cupola are shown in figure \ref{fig10} for three different values of the parameter anisotropy $\Delta$, which demonstrate a magnetocaloric response reached upon the adiabatic change of the magnetic field. Since the zero magnetization plateau is absent in magnetization curves of the spin-1/2 $XXZ$ Heisenberg pentagonal cupola [see figure \ref{fig9}~(a)], the temperature decreases to zero value when approaching zero magnetic field. It means that a magnetic compound with the magnetic structure of the spin-1/2 $XXZ$ Heisenberg pentagonal cupola should have a potential to be applied for reaching ultra-low temperatures quite similarly to the experimental realization of the spin-1/2 $XXZ$ Heisenberg triangular cupola. Moreover, if $\Delta=0$, the spin-1/2 Ising pentagonal cupola shows an enhanced magnetocaloric effect if the magnetic field is close to three values, at which the magnetization jumps in zero-temperature magnetization curve occur ($h/J=0.5, 1.5$ and $2.0$). In addition,  the spin-1/2 Heisenberg pentagonal cupola exhibits a steep change of temperature around eight values of magnetic field including zero magnetic field if $\Delta=0.5$ and $\Delta=1$.
\section{Conclusion}
\label{conclusion}
The present article deals with the ground-state phase diagram, magnetization curves and isentropes of the spin-1/2 $XXZ$ Heisenberg cupolae (triangular, square, pentagonal), which were studied using the exact numerical diagonalization. It is shown that the studied spin-1/2 $XXZ$ Heisenberg cupolae exhibit three intermediate magnetization plateaux in the limiting Ising case. Moreover, it was demonstrated that quantum ($xy$) part of the $XXZ$ exchange interaction induces the appearance of additional intermediate magnetization plateaux, which are absent in the Ising limit. While most of the novel magnetization plateaux are present at an arbitrary non-zero anisotropy parameter $\Delta>0$, there may be an exception as evidenced by a one-half plateau of a spin-1/2 $XXZ$ Heisenberg square cupola.

In the present work, the magnetocaloric effect of the spin-1/2 $XXZ$ Heisenberg cupolae was also studied. A giant magnetocaloric effect was detected in  proximity of all magnetization jumps. There is a difference between an even-numbered (square) cupola and odd-numbered (triangular and pentagonal) cupolae when considering the magnetization process. Due to the adiabatic demagnetization, the former one finally shows a small temperature rise when switching-off the magnetic field, while for the latter ones it is theoretically possible to reach the absolute zero temperature due to the absence of zero magnetization plateau. Thus, it becomes evident that the absence of zero magnetization plateau in magnetic clusters is a necessary condition to achieve a possible cooling down to the absolute zero temperature. The odd-numbered (triangular and pentagonal) cupolae with half-integer spins  fit well for this description. Hence, the experimental realization of the antiferromagnetic finite-size magnetic spin-1/2 clusters with odd total number of spins makes them promising candidates for future application in magnetic refrigeration.

\section*{Acknowledgements}
The author would like to thank to T. Verkholyak and O. Krupnitska for their valuable remarks on the first version of the manuscript. This work was financially supported by the grants of The Ministry of Education, Science, Research and Sport of the Slovak Republic under the contract Nos. VEGA 1/0531/19 and the Slovak Research and Development Agency provided under contract No. APVV-18-0197.

\newpage
\ukrainianpart

\title{Спін-1/2 XXZ модель Гайзенберґа на куполі: процес намагнічення і відповідний посилений магнетокалоричний ефект}
\author{К. Карльова }
\address{
 Інститут фізики, Факультет природничих наук, Університет імені П. Й. Шафарика, парк Ангелінум 9, Кошиці 04001, Словаччина
}

\makeukrtitle

\begin{abstract}
За допомогою методу точної діагоналізації досліджено криві намагніченості та магнетокалоричний ефект під час адіабатичної демагнетизації антиферомагнітних кластерів у формі твердих тіл Джонсона (трикутного, квадратного та п’ятикутного куполів) для різних значень обмінної анізотропії від Ізінґової до повністю ізотропної границі. Продемонстровано, що спін-1/2 XXZ Гайзенберґовий купол виявляє щонайменше одне додаткове плато намагнічення в порівнянні з відповідною йому Ізінґовою моделлю. Нове плато намагнічення поширюється на дальшу область магнітних полів зі зростанням квантової частини XXZ обмінної взаємодії за рахунок первісних плато, які присутні у границі моделі Ізінґа. Показано, що спін-1/2 XXZ модель Гайзенберґа на трикутному та п'ятикутному куполах виявляє посилений магнетокалоричний ефект поблизу нульового магнітного поля, що робить такі магнітні системи перспективними охолоджувачами до ультранизьких температур.
\keywords XXZ Гайзенберґові кластери, плато намагнічення, магнетокалоричний ефект
\end{abstract}

\end{document}